\documentclass[twocolumn,twoside,slac_two]{revtex4}
\usepackage{graphicx}
\usepackage{fancyhdr}
\newcommand{\MET}{\mbox{$E\kern-0.60em\raise0.10ex\hbox{/}_{T}$}}
\newcommand{\MetVec}{\mbox{$\vec E\kern-0.60em\raise0.10ex\hbox{/}_{T}$}}
\newcommand{\METspec}{\MET_{\mathrm{spec}}}
\newcommand{\ttbar}{t \overline{t}}
\newcommand{\llll}{\ell\ell\ell\ell}
\newcommand{\llln}{\ell\ell\ell\nu}
\newcommand{\llnn}{\ell\ell\nu \overline{\nu}}
\newcommand{\lnnn}{\ell\nu\nu \overline{\nu}}
\newcommand{\MetDeltaPhi}{\Delta\phi_{\MET,{\rm nearest}}}
\newcommand{\ppbar}{p \overline{p}}

\begin{document}

\title{Search for Standard Model Higgs Boson in H $\to$ WW Channel at CDF}

%

\author{J.~Pursley, on behalf of the CDF Collaboration}
\affiliation{Department of Physics, University of Wisconsin-Madison, Madison, WI, 53706, USA}

\begin{abstract}
We present a search for standard model (SM) Higgs boson to $WW^{(*)}$ production in 
dilepton plus missing transverse energy final states in data collected by the CDF II 
detector corresponding to 4.8 fb$^{-1}$ of integrated luminosity.
To maximize sensitivity, the multivariate discriminants used to separate signal from 
background in the opposite-sign dilepton event sample are independently optimized for 
final states with zero, one, or two or more identified jets. All significant Higgs boson 
production modes (gluon fusion, associated production with either a $W$ or $Z$ boson, and vector 
boson fusion) are considered in determining potential signal contributions. We also incorporate a 
separate analysis of the same-sign dilepton event sample which potentially contains additional 
signal events originating from associated Higgs boson production mechanisms.  Cross section 
limits relative to the combined SM predictions are presented for a range of Higgs boson
mass hypotheses between 110 and 200 GeV/$c^2$.
\end{abstract}

\maketitle

\thispagestyle{fancy}


\section{Introduction\label{sec:intro}}
The Higgs boson is introduced into the SM to explain electroweak symmetry breaking 
and the origins of particle mass.  This paper presents an analysis searching for a Higgs
boson decaying to $WW^{(*)}$, which is the dominant decay mode for $m_H>135$ GeV$/c^2$ 
\cite{Spira:1998wh}.  The small cross section of the dominant gluon fusion production 
mechanism ($\sigma_{NNLL}(gg \to H) = 0.439$ pb for $m_H=160$ GeV$/c^2$
\cite{deFlorian:2009hc,Anastasiou:2008tj,Martin:2009iq}) 
makes observation of the signal difficult within the hadron 
collider environment.  This analysis searches for $H \to WW^{(*)}\to \ell^+\ell^-\nu \overline{\nu}$, 
where $\ell$ = $e$, $\mu$, or $\tau$ with final states $e^+e^-$, $e^{\pm}\mu^{\mp}$, and 
$\mu^+\mu^-$.  In addition, a search is made for associated Higgs boson production in events with two
like-sign dileptons, with final states $e^\pm e^\pm$, $e^{\pm}\mu^{\pm}$, and $\mu^\pm\mu^\pm$.  
The data corresponds to 4.8 ${\rm fb}^{-1}$ of integrated luminosity collected 
at the Fermilab Tevatron in $\ppbar$ collisions at $\sqrt{s} = 1.96$ TeV.

\section{Event Selection\label{sec:selection}}
The CDF II detector is described in detail in Ref.~\cite{CDFII}.  The geometry of the detector
is characterized by the azimuthal angle $\phi$ and the pseudorapidity 
$\eta = -\ln [ \tan (\theta/2) ]$, where $\theta$ is the polar angle measured from 
the proton beam direction.  The transverse energy $E_T=E\sin\theta$, 
where $E$ is the energy in the calorimeter towers associated with a cluster of energy 
deposition.  Transverse momentum $p_T$ is the track momentum component transverse 
to the beam-line.  The missing transverse energy vector $\MetVec$ is defined as 
$-\sum_{i} E_T^i ~ \hat{n}_T^i$, where $\hat{n}_T^i$ is the transverse component of the 
unit vector pointing from the interaction point to the energy deposition in calorimeter tower $i$.   
This is corrected for the $p_T$ of muons, which do not deposit all of their energy in the 
calorimeter, and tracks which point to uninstrumented regions in the calorimeter.  
The missing transverse energy $\MET$ is defined as $|\MetVec|$.
Strongly interacting partons 
produced in the $\ppbar$ collision undergo fragmentation that results in highly 
collimated jets of hadronic particles.  Jet candidates are reconstructed using the calorimeter 
towers with corrections to improve the estimated energy~\cite{JES} and are required to 
have $E_T > 15$ GeV and $|\eta| < 2.5$.

The $\llnn$ candidates are selected from two opposite-sign (OS) leptons.  At least one lepton 
is required to satisfy the real-time trigger selection and have $p_T > 20$ GeV$/c$.  This 
requirement is loosened to $10$ GeV$/c$ for the second lepton to increase Higgs boson kinematic 
acceptance.  The $z$-positions of the leptons in a candidate at the point of closest approach 
to the beam-line are required to be within 4~cm of each other to reduce backgrounds from
overlapping $\ppbar$ collisions.

There are several sources of background with the same or a similar final state: 
$WW\rightarrow\llnn$, $ZZ\rightarrow\llnn$ or $\llll$, $WZ\rightarrow\llln$ or $\lnnn$, 
$\ttbar\rightarrow b \overline{b}\llnn$, Drell-Yan (DY) $Z\rightarrow\ell\ell$ where the 
measured large $\MET$ is due to resolution tails, $W\gamma$ where the photon converts to an 
$e^-e^+$ pair, and $W$+jets where a jet is misidentified as a lepton.  To reduce the 
contribution of $W\gamma$ and $W$+jets, all leptons must be isolated with requirements on both
the surrounding $E_T$ in the calorimeters and $p_T$ of surrounding tracks.  Candidates are 
further required to have a dilepton invariant mass $M_{\ell\ell} > 16$ GeV/$c^2$ to suppress 
$W\gamma$, and exactly 2 leptons to suppress contributions with additional leptons.

To reduce the DY background, we require $\METspec>$~25 GeV (reduced to
$\METspec>$~15 GeV for electron-muon events where DY background is
inherently smaller), where 
\[
     \METspec \equiv \left\{ 
     \begin{array}{ll} 
     \MET                       & \mbox{ if } \MetDeltaPhi > \frac{\pi}{2} \\
     \MET\sin({\MetDeltaPhi})   & \mbox{ if } \MetDeltaPhi < \frac{\pi}{2} \\
     \end{array} \right.
\]
The $\MetDeltaPhi$ is the azimuthal separation between the $\vec{\MET}$ and the 
momentum vector of the nearest lepton or jet candidate.  With this definition, we 
require the missing energy transverse to the nearest lepton or jet in the event 
to be greater than the minimum threshold if $\vec{\MET}$ points along that object.
Thus a mismeasurement of the energy of one lepton or jet candidate
would not allow the event to enter the sample.

We consider individually final states with zero, one, or two or more jet candidates
in the event.  This allows us to tune our multivariate discriminants on the different 
relative sizes of signal and background contributions within each jet multiplicity bin.  
The dominant background in the zero jet bin is $WW$, while in the one jet bin
the background from DY is of a similar size as the $WW$ background.
The dominant background in the two or more jets bin is $\ttbar$ production.  To
suppress the $\ttbar$ contribution, all events with two or more jets containing 
one or more jets with a tight, secondary vertex $b$-tag are rejected.
After all selection requirements we observe (summing over all jet multiplicities) a 
total of 1467 candidate events compared against an expectation of $1479\pm141$ 
background events and $25\pm3$ signal events for a SM Higgs boson with a mass 
of 165~GeV/$c^2$.

\subsection{Event Selection for Same Sign Analysis\label{subsec:ss}}
Like-sign, or same-sign (SS), dileptons occur in $VH \to VWW$ production when the 
vector boson ($Z$  or $W$) and one of the $W$ bosons from the Higgs boson decays leptonically.  
The dominant backgrounds to this search are from the charge mismeasurement of a real lepton,
photon conversion to $e^-e^+$, and misidentification of a jet as a lepton.

The event selection for OS events is modified to select two like-sign leptons and reduce 
potential backgrounds.  The SS channel does not use forward electron candidates, which have 
a high rate of charge mismeasurement.  To reduce the numbers of conversion
electrons and jets misidentified as leptons, the $p_T$ requirement for the second 
lepton is increased from 10 to 20 GeV$/c$.  As the decay of the third 
boson most often results in the production of additional jets, we also require one or 
more jet candidates in the final state.  Like-sign $VH$ events tend to have lower values of 
$\MET$ because the neutrinos are not necessarily back-to-back.  
To increase the signal acceptance, no $\METspec$ requirement is made.
After all selection requirements we observe a total of 64 candidate events compared 
against an expectation of $62\pm11$ background events and $1.7\pm0.2$ signal 
events for a SM Higgs boson with a mass of 165~GeV/$c^2$.

\section{Data Modeling\label{sec:mc}}
The geometric and kinematic acceptance for the $WW$, $WZ$, $ZZ$, $W\gamma$, DY, 
$\ttbar$, and all of the signal processes ($gg \to H$, $WH$, $ZH$, VBF) 
are determined using a Monte Carlo calculation of the collision followed by a 
{\sc geant3}-based simulation of the CDF II detector response \cite{GEANT}.  The Monte Carlo 
generator used for $WW$ is {\sc mc@nlo}~\cite{Frixione:2002ik}, while for $WZ$, $ZZ$, DY, 
$\ttbar$, and the signal processes {\sc pythia}~\cite{Sjostrand:2006za} is used, and $W\gamma$ 
is modeled with the generator described in Ref.~\cite{WgXsec}. We use the {\sc cteq5l} 
\cite{Lai:1999wy} parton distribution functions (PDF) to model the momentum distribution 
of the initial-state partons. 

The expected yields of background processes are normalized to cross sections 
calculated at partial next-to next-to-leading order 
($\ttbar$~\cite{ttbarXsec}), next-to-leading order ($WW$, $WZ$, and $ZZ$~\cite{DiboXsec}), 
or leading-order with estimated higher-order corrections ($W\gamma$~\cite{WgXsec} and 
DY~\cite{DY}).  The gluon fusion cross section has been calculated to next-to 
next-to-leading logarithmic accuracy~\cite{deFlorian:2009hc,Anastasiou:2008tj,Martin:2009iq}.
Associated production 
cross sections at next-to next-to-leading order (NNLO) are obtained from Ref.~\cite{WHZH}. 
VBF cross sections at next-to-leading order (NLO) are obtained from Ref.~\cite{ref:tev4lhc}.

The background from $W$+jets is estimated from a sample of data events with one identified lepton 
and one jet that fulfills loose isolation requirements and contains a track or energy cluster 
similar to those required for lepton identification. The contribution of each event to the 
total yield is scaled by the probability that such a jet is identified as a lepton. This probability 
is determined from multijet events collected in independent jet-triggered data samples.
A correction is applied for the small real lepton contribution using single $W$ and $Z$ boson 
Monte Carlo simulation.

\section{Cross Checks\label{sec:control}}
Several control regions are constructed to validate different aspects of the data 
modeling.  The sample sizes of the control regions are designed to be large 
enough to give statistically meaningful tests.  These control regions, described
below, show good agreement between the background model predictions and the
observed distributions in data.

One of the most fundamental control regions is the Drell-Yan sample selected 
using the default OS candidate selection in Sec.~\ref{sec:selection}
with a reversed $\MET < 25$ GeV (15 GeV in the case of $e\mu$ events)
requirement and the additional requirement that the dilepton invariant mass 
fall in the $Z$ mass region.  This control sample 
provides a high statistics test of acceptance, lepton identification efficiency, 
and trigger efficiency calculations.

A $W$+jets control sample is constructed from events which satisfy the SS candidate 
selection in Sec.~\ref{subsec:ss} but have zero jet candidates.  This control sample 
contains contributions primarily from jets misidentified as leptons, with additional 
contributions from conversion electrons and charge mismeasurement of a real lepton,
and checks the modeling of misidentified particles.

A $\ttbar$ control sample is constructed using the default OS candidate selection in 
Sec.~\ref{sec:selection} with a requirement of two or more jets in the event, and 
the $b$-tag veto requirement reversed such that only events with one or more jets 
with a $b$-tag are accepted.

A second DY control sample tests the $\MET$ model, and is constructed using 
the default OS candidate selection in Sec.~\ref{sec:selection} with the exception 
of requiring 15 $<\METspec<$ 25 GeV instead of $\METspec>$ 25 GeV 
(in the case of $e\mu$ events, 10 $<\METspec<$ 15 GeV instead of $\METspec>$ 15 GeV).  
No requirements are made on the jet multiplicity or the dilepton invariant mass.

\section{Analysis Technique\label{sec:technique}}
For each search channel, a NeuroBayes$^\textrm{\textregistered}$~\cite{Feindt:2006pm}
neural network is trained on a weighted combination of signal and background events from 
Monte Carlo independently for each Higgs boson mass hypothesis (14 total). Each neural 
network has three layers consisting of input nodes, hidden nodes, and one output node. 
Once the neural network is trained, templates are created for all signal and background 
processes. For the OS zero and one jet channels, ``high S/B'' and ``low S/B'' templates 
based on the signal-to-background ratio of the different dilepton combinations are 
considered separately.  The templates are used as the final discriminant in calculating 
the 95\% credibility level (C.L.) limits.

The OS zero jet channel uses 5 input variables. The inputs are: the $\Delta R$ between the leptons, 
the $\Delta \phi$ between the leptons, the scalar sum of the leptons' $p_T$ and the $\MET$, 
the likelihood ratio for Higgs to $WW$ production ($LR_{HWW}$), and the likelihood ratio for $WW$ 
production ($LR_{WW}$). The likelihood ratios are determined using a matrix element method which 
calculates the probability for each event to have been produced by several relevant SM processes.

The OS one jet channel uses 8 input variables. The inputs are: the $M_{\ell\ell}$, 
the transverse mass of the sum of the leptons' 4-momenta and ($\MET$, MetX, MetY, 0) where 
MetX and MetY are the $x$ and $y$ components of the $\vec{\MET}$, the $\Delta R$ 
between the leptons, the scalar sum of the leptons' $p_T$ and jets' $E_T$ and the $\MET$ ($H_T$), 
the $\METspec$, the leading lepton $p_T$, the subleading lepton $p_T$, and the
leading lepton energy.

The OS two or more jets channel uses 8 input variables. The inputs are: the $M_{\ell\ell}$, the
leading lepton $p_T$, the subleading lepton $p_T$, the $H_T$, the $\Delta R$ between the 
leptons, the $\Delta\phi$ between the leptons, the $\Delta\phi$ between the vector sum of the 
leptons' $p_T$ and the $\MET$, and the vector sum of the first and second jet $E_T$.

The SS one or more jets channel uses 13 input variables.  The inputs are: the $M_{\ell\ell}$, 
the leading lepton energy, the leading lepton $p_T$, the subleading lepton $p_T$, the $H_T$, 
the leading jet $E_T$, the $\Delta\phi$ between the leptons, the vector sum of all jets' $E_T$, 
the vector sum of leptons' $p_T$, the $\MET$, the $\MET$ 
significance ($\MET$/$\sqrt{\Sigma E_T}$), the $\METspec$, and the $\MetDeltaPhi$.

\begin{figure*}
\includegraphics[width=80mm]{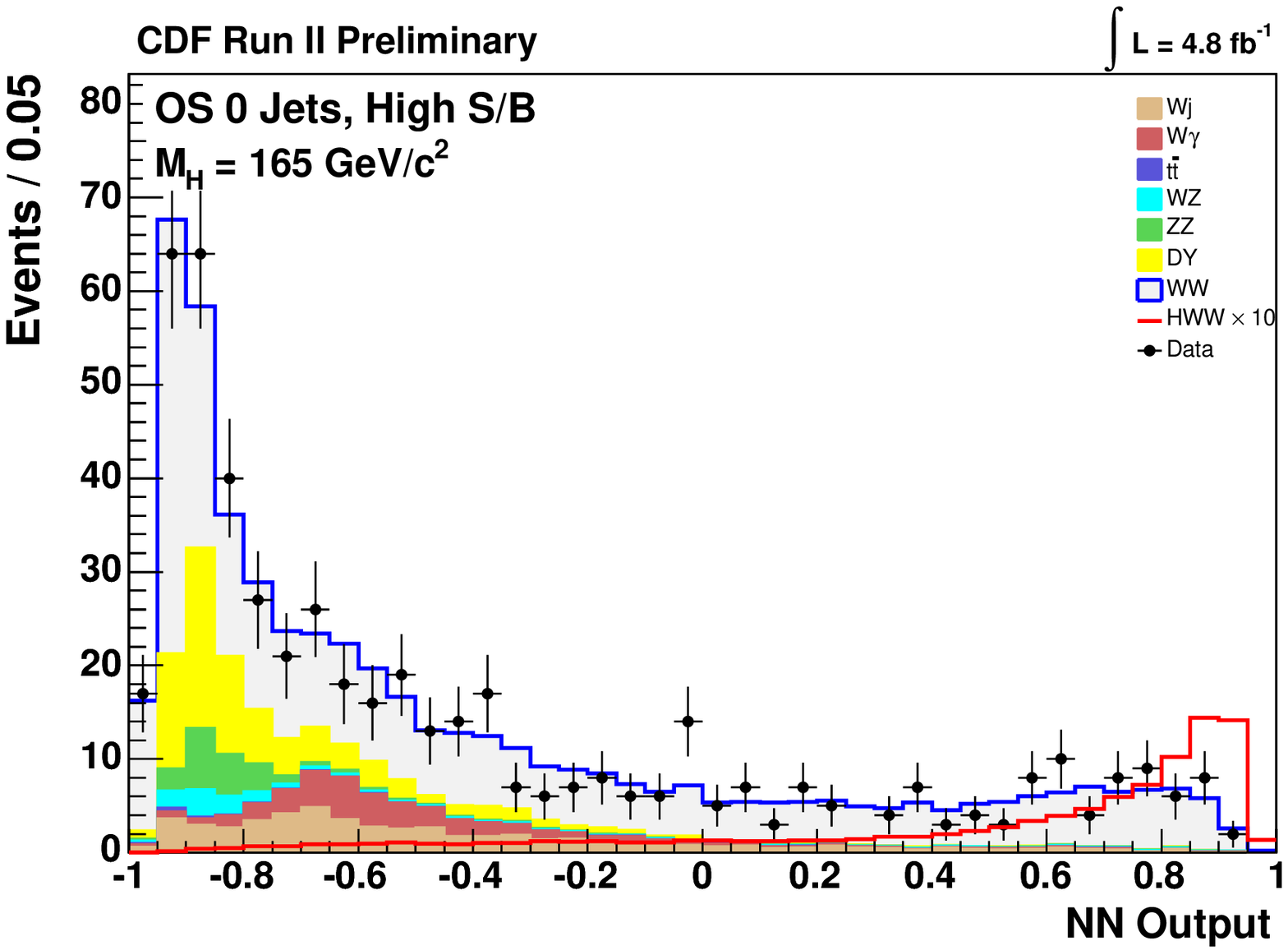}
\includegraphics[width=80mm]{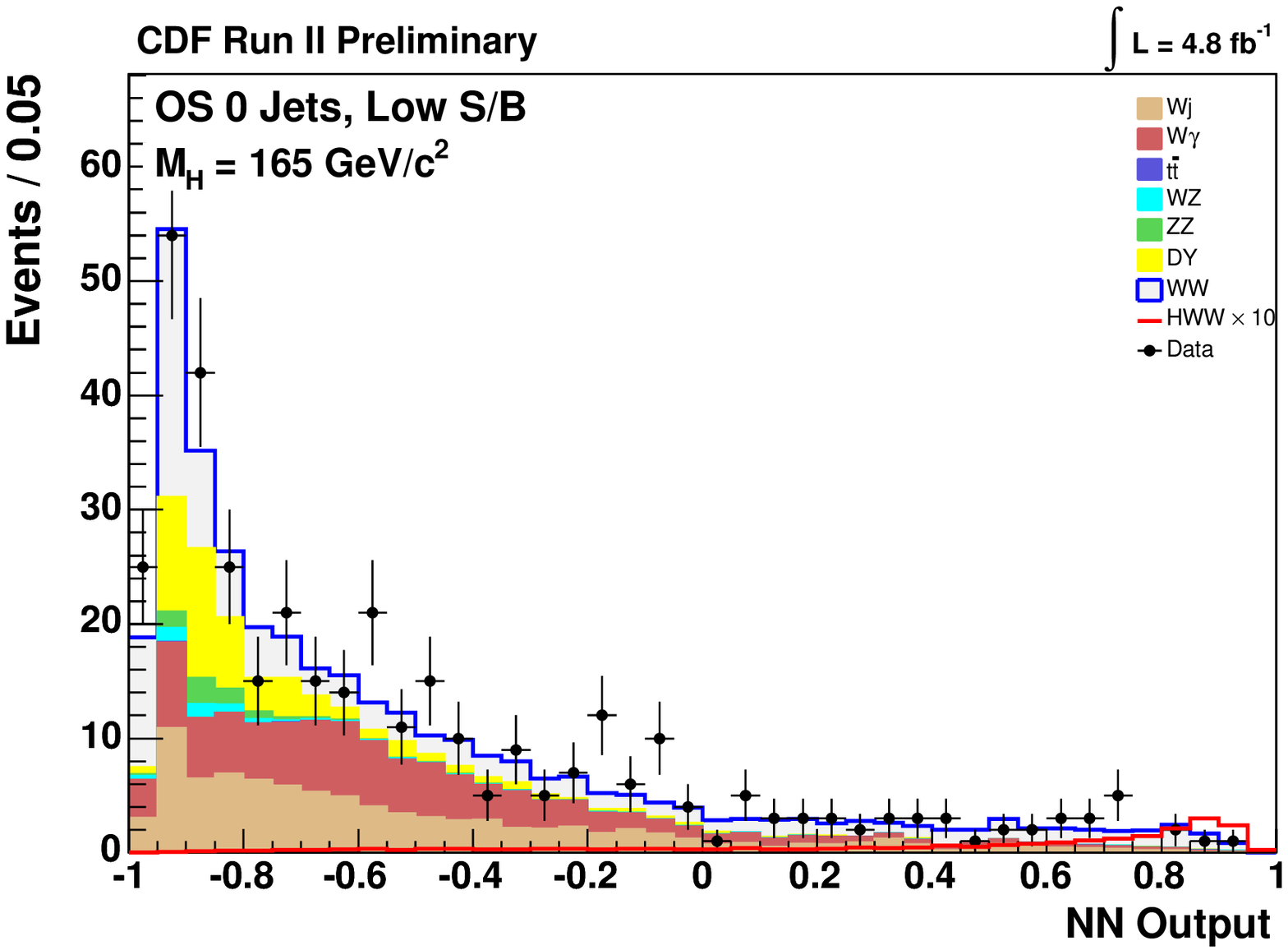}
\includegraphics[width=80mm]{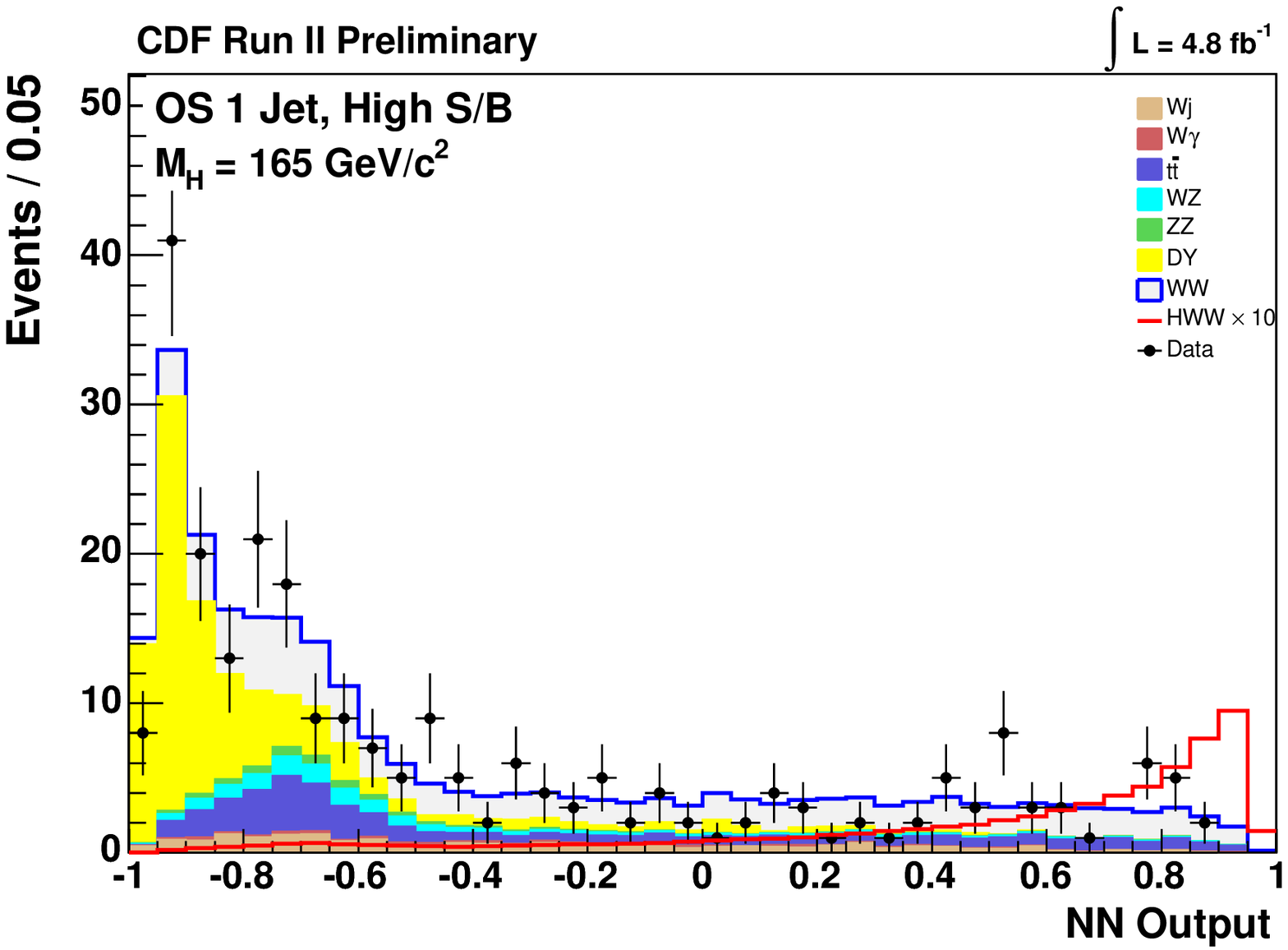}
\includegraphics[width=80mm]{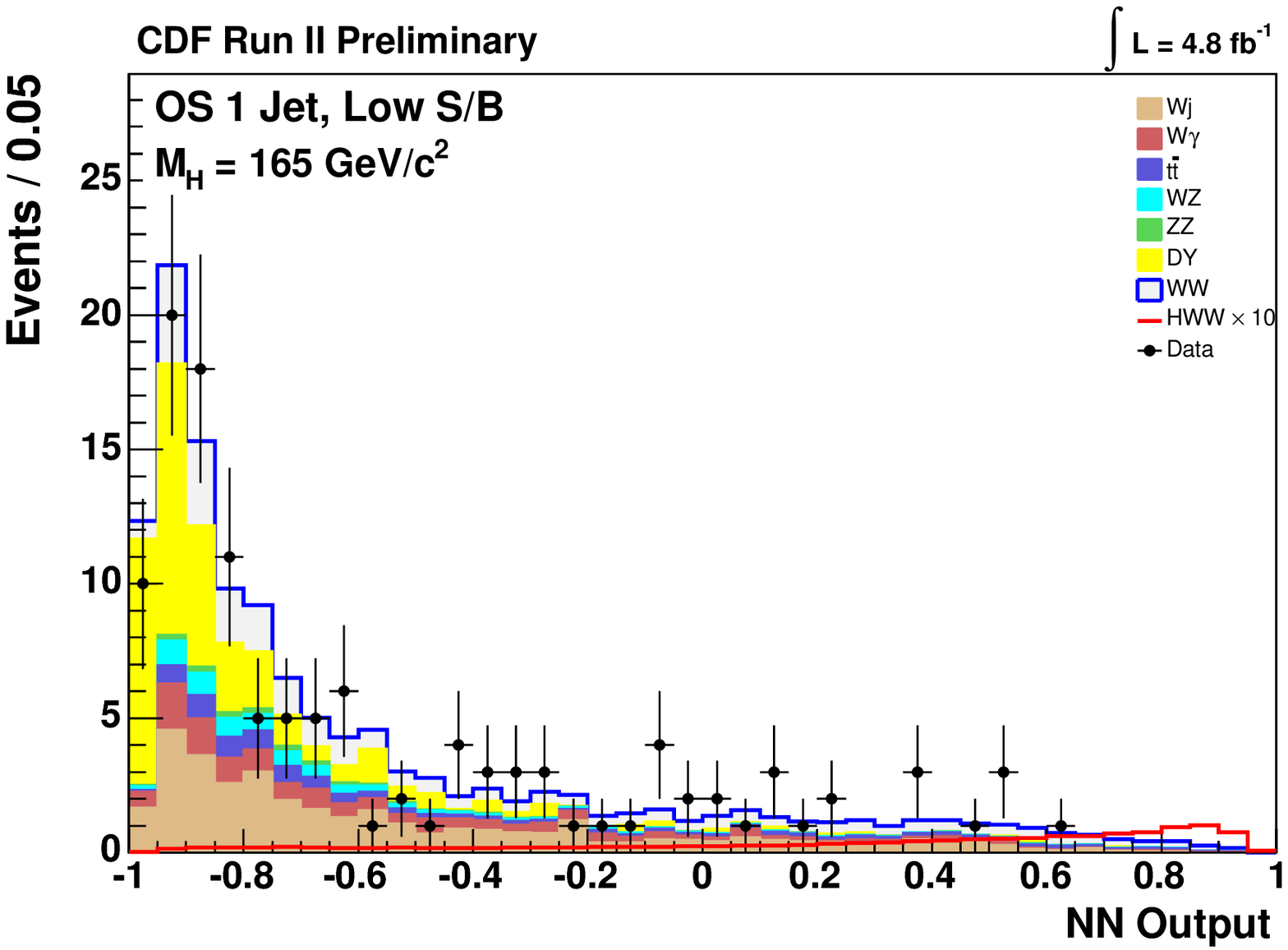}
\includegraphics[width=80mm]{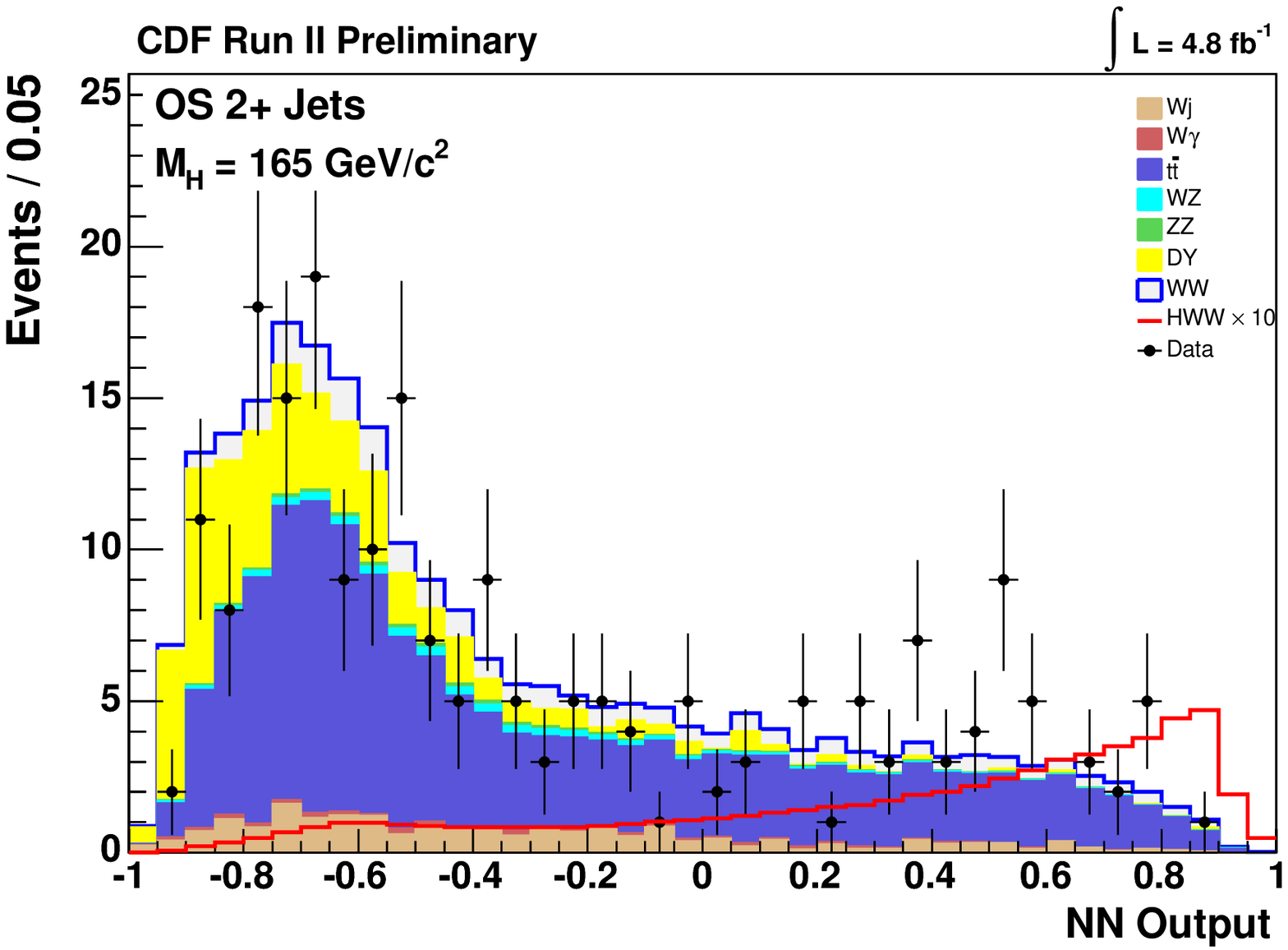}
\includegraphics[width=80mm]{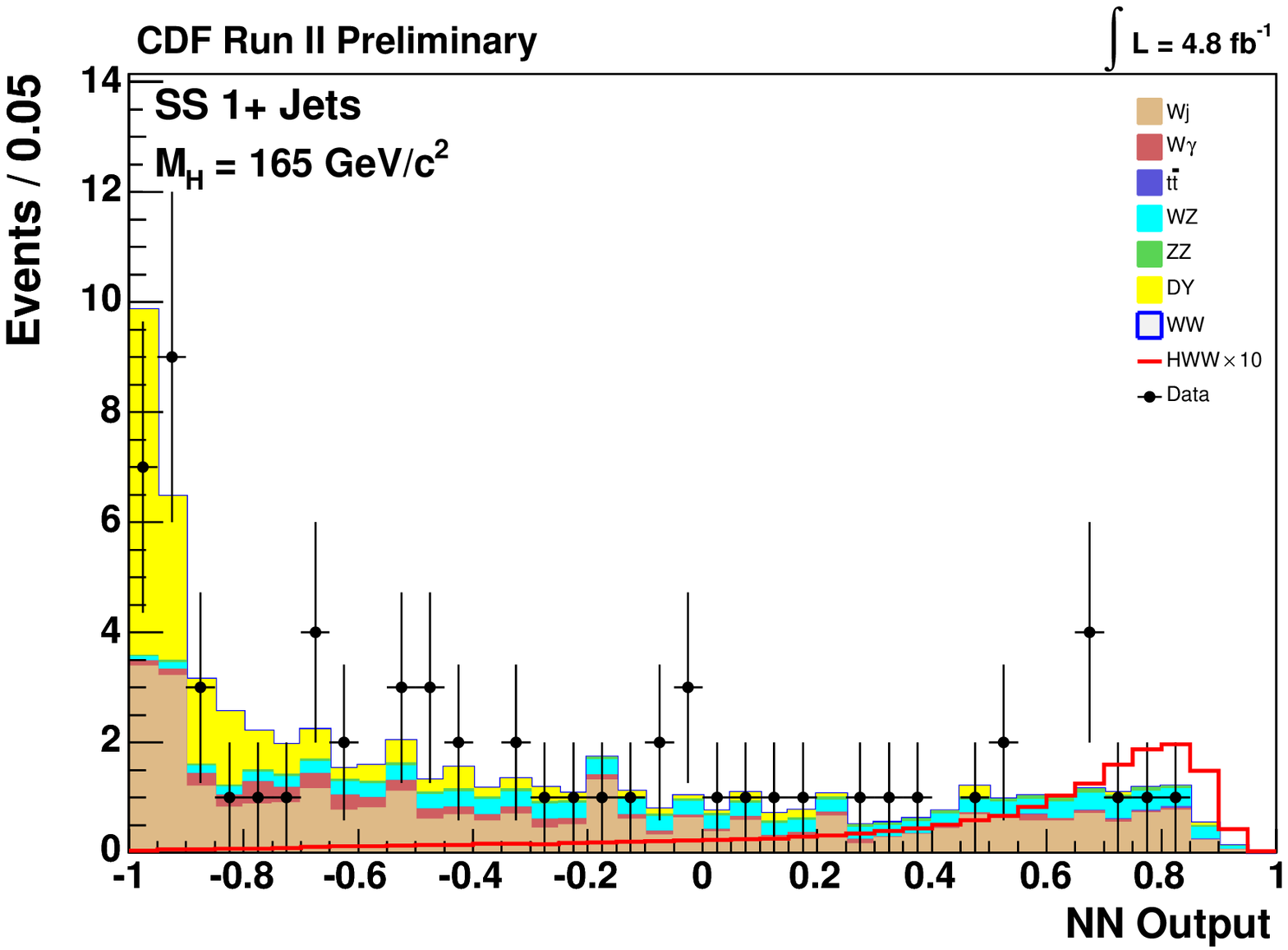}
\caption{Neural network outputs for each channel (OS zero and one jet for both high
  and low S/B, OS two or more jets, and SS one or more jets) at $m_H = 165$ GeV$/c^2$.  
  The Higgs boson signal is shown at 10 times the SM prediction.\label{fig1}}
\end{figure*}

Neural network outputs are shown in Figure~\ref{fig1} for each channel at a Higgs boson
mass of 165 GeV$/c^2$.  Plots of the input variables and neural network outputs for all Higgs 
boson masses are available in Ref.~\cite{HWWweb}.

%


\section{Systematics\label{sec:syst}}
Systematic uncertainties associated with the Monte Carlo simulation affect the  
signal, $WW$, $WZ$, $ZZ$, $W\gamma$, DY, and $\ttbar$ acceptances taken from 
the simulated event samples. Uncertainties originating from lepton selection and 
trigger efficiency measurements are propagated through the acceptance calculation, 
giving uncertainties typically around 2\%, but as high as 7\% for the different 
signal and background processes.  
For the SS channel, there is an additional systematic for the simulated 
backgrounds which enter the sample only when a lepton's charge is mismeasured 
($WW$, $\ttbar$, and DY) which is taken as half the difference between 
the mismeasurement rates found in DY events and those found in Monte Carlo.

We also assign an acceptance uncertainty due to potential contributions from 
higher-order effects.  In the case of $WW$ we take half of the difference 
between the leading-order ({\sc pythia}-based~\cite{Sjostrand:2006za}) and 
next-to-leading order ({\sc mc@nlo}~\cite{Frixione:2002ik}) acceptances.  
The other processes are only simulated at leading-order and for these modes 
we assign the full difference observed in $WW$, leading to a 10\% uncertainty.
The largest uncertainty on the Drell-Yan process originates from modeling of
the fake $\MET$ in these events.  Based on discrepancies in the DY $\MET$
control region, we assign a 20\% uncertainty in the 
zero jet channel and 25\% uncertainties in the one and two or more jets channels.
For processes that produce no final state jets at leading-order ($WW$, 
$W\gamma$, and DY), we also assign a jet modeling systematic to account for 
uncertainites in the Monte Carlo modeling of the $p_T$ boost and additional 
jet production from initial and final state radiation.  The uncertainties for 
these effects are anti-correlated between jet channels since the effect of 
extra jet production is to move events from one channel to another.  For Monte 
Carlo samples not currently generated over the entire data run range, we take 
additional uncertainties on the acceptance for the corresponding signal or 
background process.  This uncertainty comes from the observed difference 
in the leading-order $WW$ acceptance for the corresponding partial run range 
versus that for the full run range.  The acceptance variations due to PDF
model uncertainties is assessed to be on the order of 2\% using the 20 
pairs of PDF sets described in Ref.~\cite{Kretzer:2003it}. 

For the $W\gamma$ background contribution, there is an additional uncertainty of 
20\% from the detector material description and conversion veto efficiency. The 
systematic uncertainty on the $W$+jets background contribution is determined from 
differences in the measured probability that a jet is identified as a lepton for 
jets collected using different jet $E_T$ trigger thresholds. These variations 
correspond to changing the parton composition of the jets and the relative amount 
of contamination from real leptons. 
  
The cross section calculation uncertainties are 6\% on diboson production, 
10\% on $\ttbar$ and $W\gamma$ production, and 5\% on Drell-Yan production.  All
signal and background estimates obtained from simulation have an additional 
6\% uncertainty originating from the luminosity measurement~\cite{Acosta:2002hx}.

Most systematic uncertainties on the signal processes are assessed using the 
same techniques described for the background processes.  Uncertainties on the 
theoretical cross sections vary by Higgs boson production mechanism.  
Associated production cross sections are known to NNLO, so the theoretical 
uncertainty on these cross sections is small, less than 5\%~\cite{ref:tev4lhc}.  
VBF production is known only to NLO, so the residual 
theoretical uncertainty is higher, on the order of 10\%~\cite{ref:tev4lhc}.  
Gluon fusion is a QCD process; thus the theoretical uncertainty is still 
significant even though it is known to NNLO.  We use the 
{\sc hnnlo} program~\cite{Catani:2007vq,Grazzini:2008tf}, which computes 
the theoretical cross section for $gg \to H$ production at the Tevatron to 
NNLO, to assign theoretical uncertainties due to changing the renormalization 
and factorization scales (11\%) and the gluon PDF model (5\%).  
We again use the {\sc hnnlo} program to evaluate acceptance uncertainties for 
$gg \to H$ production for both changes in scale and the PDF model.  To evaluate the 
change in acceptance, we reweight the Higgs boson $p_T$ and $\eta$ distributions 
obtained from {\sc pythia} to match the {\sc hnnlo} calculations.  Reweighting 
events based on Higgs boson $p_T$ changes the relative acceptances for the jet 
channels, while reweighting events based on 
Higgs boson $\eta$ primarily affects the lepton acceptance.

\section{Results\label{sec:results}}
To determine the sensitivity we construct a binned likelihood using the six neural 
network templates shown in Figure~\ref{fig1}: two templates each for the OS zero and one
jet channels (high S/B and low S/B), one template for the OS two or more jets channel,
and one template for the SS one or more jets channel.  All the signal and background normalizations are 
allowed to float, but ratios of signal and background contributions are constrained to 
their expectations with a set of Gaussian constraints determined from the 
systematic uncertainties. The total signal yield is allowed to float.

\begin{figure*}
\includegraphics[width=150mm]{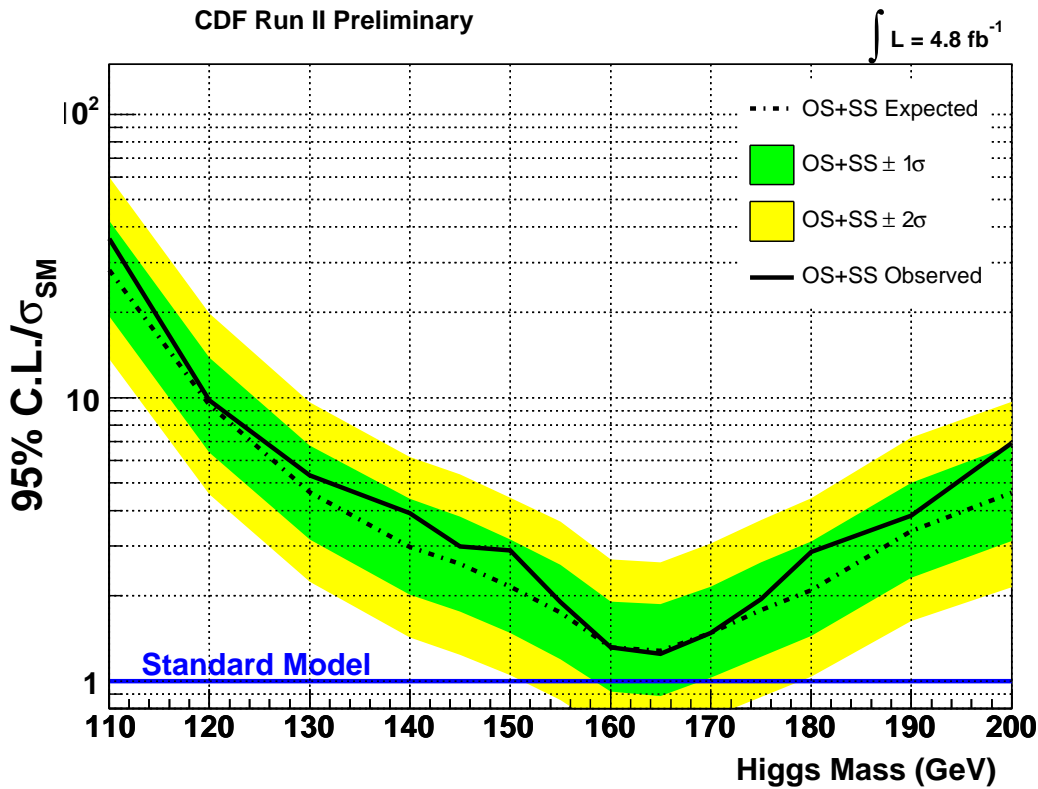}
\caption{The ratio of expected and observed 95\% C.L. production cross section 
	limits to SM predictions for 14 different Higgs boson masses for the combination 
	of all opposite-sign and same-sign $H \to WW^{(*)}$ decay channels.  Error bands are shown
	on the expected limit, while the observed limit is shown as a solid line.\label{fig2}}
\end{figure*}

The 95\% C.L. limits are determined with a set of 50,000 Monte Carlo 
background-only experiments based on expected yields varied within the assigned systematics.  
For each experiment a test statistic is formed from the difference in the likelihood 
value for the background-only model versus that for the signal plus background model.  
The median expected 95\% C.L. limit at a Higgs mass of 165 GeV/$c^2$ is 
$1.3^{+0.6}_{-0.4}$ times 
the SM prediction, while the observed limit is 1.3 times the SM prediction.  
Results for 14 values of $m_H$ are shown in Table~\ref{tab1} and Figure~\ref{fig2}.

%
\begin{table*}
\caption{The ratio of expected and observed 95\% C.L. production cross section 
	limits to SM predictions for 14 different Higgs boson masses for the combination 
	of all opposite-sign and same-sign $H \to WW^{(*)}$ decay channels.\label{tab1}}
\begin{tabular}{|l|r|r|r|r|r|r|r|r|r|r|r|r|r|r|}
\hline
$m_H~({\rm GeV}/c^2)=$   &         110 &         120 &         130 &         140 &         145 &         150 &         155 &         160 &         165 &         170 &         175 &         180 &         190 &         200 \\
\hline
     $-2\sigma/\sigma_{SM}$ &       13.70 &        4.55 &        2.24 &        1.42 &        1.24 &        1.05 &        0.86 &        0.66 &        0.64 &        0.74 &        0.88 &        1.04 &        1.63 &        2.14 \\
     $-1\sigma/\sigma_{SM}$ &       19.38 &        6.41 &        3.15 &        2.01 &        1.76 &        1.48 &        1.19 &        0.92 &        0.89 &        1.03 &        1.22 &        1.45 &        2.31 &        3.12 \\
   $\bf Median/\sigma_{SM}$ & \bf   28.23 & \bf    9.45 & \bf    4.63 & \bf    2.97 & \bf    2.59 & \bf    2.15 & \bf    1.74 & \bf    1.32 & \bf    1.28 & \bf    1.48 & \bf    1.78 & \bf    2.09 & \bf    3.39 & \bf    4.61 \\
     $+1\sigma/\sigma_{SM}$ &       41.99 &       13.85 &        6.79 &        4.39 &        3.81 &        3.16 &        2.57 &        1.91 &        1.87 &        2.16 &        2.62 &        3.11 &        5.01 &        6.81 \\
     $+2\sigma/\sigma_{SM}$ &       60.07 &       19.83 &        9.61 &        6.18 &        5.35 &        4.44 &        3.65 &        2.68 &        2.63 &        3.06 &        3.70 &        4.42 &        7.23 &        9.68 \\
\hline
 $\bf Observed/\sigma_{SM}$ & \bf   36.44 & \bf    9.78 & \bf    5.33 & \bf    3.92 & \bf    2.99 & \bf    2.89 & \bf    1.90 & \bf    1.31 & \bf    1.25 & \bf    1.48 & \bf    1.95 & \bf    2.86 & \bf    3.84 & \bf    6.93 \\
\hline
\end{tabular}
\end{table*}

\begin{acknowledgments}
We thank the Fermilab staff and the technical staffs of the participating institutions 
for their vital contributions. This work was supported by the U.S. Department of Energy 
and National Science Foundation; the Italian Istituto Nazionale di Fisica Nucleare; the 
Ministry of Education, Culture, Sports, Science and Technology of Japan; the Natural 
Sciences and Engineering Research Council of Canada; the National Science Council of 
the Republic of China; the Swiss National Science Foundation; the A.P. Sloan Foundation; 
the Bundesministerium f\"ur Bildung und Forschung, Germany; the World Class University 
Program, the National Research Foundation of Korea; the Science and Technology Facilities 
Council and the Royal Society, UK; the Institut National de Physique Nucleaire et Physique 
des Particules/CNRS; the Russian Foundation for Basic Research; the Ministerio de Ciencia 
e Innovaci\'{o}n, and Programa Consolider-Ingenio 2010, Spain; the Slovak R\&D Agency; 
and the Academy of Finland. 
\end{acknowledgments}

\bigskip 

\end{document}